\documentclass[conference,A4paper]{IEEEtran}
\IEEEoverridecommandlockouts
\usepackage{cite}
\usepackage{amsmath,amssymb,amsfonts}
\usepackage{algorithmic}
\usepackage[caption=false,font=footnotesize]{subfig}
\usepackage{graphicx}
\usepackage{textcomp}
\usepackage{xcolor}
\usepackage{siunitx}
\usepackage{url}
\usepackage{tabularx}
\usepackage{multirow}
\usepackage{booktabs}
\usepackage{color,soul}
\usepackage{makecell}
\usepackage[colorlinks=true, allcolors=black]{hyperref}
\usepackage{array}

\newcommand{\cmark}{\checkmark}

\begin{document}

\title{A Modular Cyber Range Platform for Smart Energy Systems
\thanks{
This work was supported by the ACTING Project, Co-funded by the European Union through the European Defence Fund (EDF) and in part by the European Union’s Horizon 2020 research and innovation programme under Grant Agreement No 739551 (KIOS CoE-TEAMING) and from the Republic of Cyprus through the Deputy Ministry of Research, Innovation and Digital Policy. Views and opinions expressed are however those of the author(s) only and do not necessarily reflect those of the European Union or European Commission. Neither the European Union nor the granting authority can be held responsible for them.
}
}

\author{\IEEEauthorblockN{\IEEEauthorrefmark{1}Vasilis Ieropoulos, 
\IEEEauthorrefmark{1}Theofanis~Eleftheriadis,
\IEEEauthorrefmark{1}Kyriakos~Christou,
\IEEEauthorrefmark{1}Philippos~Isaia, 
\IEEEauthorrefmark{1}Maria~Michalopoulou,\\
\IEEEauthorrefmark{1}\IEEEauthorrefmark{2}Angelos K. Marnerides,
\IEEEauthorrefmark{1}Christos~Laoudias,
\IEEEauthorrefmark{1}\IEEEauthorrefmark{2}Maria~K.~Michael 
}
\IEEEauthorblockA{\IEEEauthorrefmark{1}KIOS Research and Innovation Center of Excellence, University of Cyprus, Cyprus}
\IEEEauthorblockA{\IEEEauthorrefmark{2}Department of Electrical and Computer Engineering, University of Cyprus, Cyprus}
}

\maketitle

\begin{abstract}
Modern critical infrastructure requires realistic training tools that bridge the gap between digital vulnerabilities and physical consequences. This paper introduces a lightweight, modular Cyber Range (CR) designed to simulate Information Technology (IT) and Operational Technology (OT) environments with high fidelity.
Using the FastAPI framework as backend and Quick Emulator (QEMU)/Kernel-based Virtual Machine (KVM) virtualisation, the platform employs a YAML-based engine to automatically deploy complex networks on standard hardware with minimal overhead. To demonstrate its capabilities, we present a Photovoltaic (PV) Plant scenario set in a fictional smart city. The scenario follows a multi-stage attack that begins with the compromise of a home Internet of Things (IoT) device and escalates into a major power outage at a military base. The proposed CR, natively supports industrial protocols such as Modbus/TCP and Message Queuing Telemetry Transport (MQTT) and includes a "Green Team" framework for evaluating human performance during crisis situations. This paper describes the system architecture, automated setup, protocol-simulation capabilities, and scenario design.
\end{abstract}


\section{Introduction}
\label{sec:intro}

The increasing sophistication of cyber threats has made cyber readiness a critical requirement across the public, private, and critical infrastructure sectors~\cite{admass2024cyber}. Smart cities increasingly depend on interconnected energy, water, transportation, healthcare, telecommunication, cloud, and IoT systems. However, this connectivity expands the attack surface and introduces dependencies between digital services and physical processes. Cybersecurity is no longer only a matter of protecting data; it is a prerequisite for public safety, economic stability, and operational continuity.

Modern adversaries conduct persistent, multi-vector operations targeting not only information systems but also operational technologies, supply chains, and interconnected infrastructure. Real-world incidents, such as attacks on power systems, ransomware affecting fuel distribution, and disruptions to healthcare services, demonstrate how cyber incidents can propagate across essential services \cite{yu2023survey}. Consequently, organisations require realistic environments for training, testing, and evaluating incident response. Cyber Ranges (CRs) provide controlled, reproducible environments for defenders, operators, and decision-makers to conduct exercises and validate procedures \cite{yamin2020cyber}. These platforms are especially relevant for cyber-physical environments where testing on operational infrastructure is unsafe or impractical. A \textbf{Cyber-Physical System (CPS)} integrates computation, networking, and physical processes. The type of CR examined in this research is specifically designed for CPS, in which both cyber actions and their physical consequences must be simulated in a realistic environment. This capability is termed high-fidelity simulation, meaning the platform accurately models both low-level network protocols (e.g., Modbus/TCP) and the operational impact on physical processes (e.g., changes in power output).

Recent work has highlighted that CRs are not merely technical testbeds, but complex socio-technical ecosystems that combine infrastructure resources, orchestration mechanisms, learning services, assessment processes, and human-in-the-loop activities~\cite{kampourakis2026ontology}. This broader understanding is important for platforms targeting critical infrastructure, where realistic training requires both technical fidelity and meaningful representation of operational roles, workflows, and decision-making processes.

Despite their value, standalone CRs often lack the scale, diversity, and interoperability needed to represent modern digital ecosystems~\cite{Park2022}. The absence of common architectural principles and standardised representations also makes it difficult to compare, integrate, and evaluate different CR platforms. To address this gap, recent research has proposed reference architectures that organise CR components across structural, functional, and informational dimensions, providing a more systematic basis for CR design, development, and evaluation~\cite{kampourakis2024reference}. In parallel, ontology-based approaches have been introduced to semantically represent CR concepts, roles, resources, scenarios, and capabilities, supporting more consistent reasoning about CR ecosystems and their configurations~\cite{kampourakis2026ontology}.

Federated CRs overcome the limitations of isolated platforms \cite{Virag2021} by interconnecting independent ranges to share resources and jointly execute coordinated scenarios \cite{ECSO2020}. This approach is essential for cross-sector, cyber-physical scenarios where a single platform cannot reproduce the complex dependencies among energy, communication, transport, cloud, and operational control systems. Consequently, federation amplifies the need for clear architectural models, interoperable descriptions, and systematic evaluation methods.

Beyond technical fidelity, effective cyber training must address human and organisational factors across all roles, from engineers to managers. Training environments must support protocol experimentation and attack simulation while assessing awareness, coordination, decision-making, and response procedures. Recent work on explainable CR evaluation emphasises structured assessment criteria covering technical fidelity, training capabilities, scalability, usability, and mission relevance \cite{Kampourakis2025LLMAssistedAF}. These structured perspectives ensure that CRs provide measurable improvements in overall cyber readiness.

A well-designed CR evaluates both technical and human dimensions through distinct roles: the \textbf{Red Team} emulates adversaries to generate realistic attack scenarios \cite{YULIANTO2025100077}, the \textbf{Blue Team} monitors and defends to demonstrate organisational detection capabilities~\cite{10.1007/978-3-030-01168-0_26}; and the \textbf{Green Team} represents operational personnel to measure resilience against anomalies and social engineering~\cite{PRUMMER2025104206}. This interaction provides a holistic readiness assessment framework.

The interaction among the Red, Blue, and Green Teams provides a holistic evaluation framework that assesses technical exploit validation alongside human resilience, communication, and decision-making under stress. Within this context, this paper introduces a modular CR platform designed for realistic cyber-physical training and experimentation. The platform supports the automated deployment of complex Information and Communication Technology (ICT) and OT scenarios involving interconnected systems, industrial protocols, and cascading operational effects. The following sections describe the system architecture, orchestration engine, protocol-simulation capabilities, and a representative smart-energy scenario, while empirical validation of team-based evaluation is reserved for future work.

The remainder of the paper is organised as follows. Section~\ref{sec:RelatedWork} reviews related work. Section~\ref{sec:Architecture} presents the proposed CR system architecture and scenario design. Section~\ref{sec:Evaluation} details the evaluation of the system's usability and functionality through the PV Plant scenario. Section~\ref{sec:Conclusions_FutureWork} concludes and outlines future directions.

\section{Related Work}
\label{sec:RelatedWork}

CRs vary across academic, government, and commercial domains, each prioritizing different levels of openness, automation, and specificity. In academia, the open-source \textbf{KYPO}~\cite{9637180} leverages infrastructure-as-code principles via YAML for automated deployment but focuses heavily on IT environments. Similarly, \textbf{CyTrONE}~\cite{10.1145/3011077.3011087} automates range instantiation using its CyRIS component but lacks native industrial protocol support or human-centric assessment frameworks. 

Government-supported infrastructure like the \textbf{AIT CR}~\cite{10.1145/3424954.3424959} delivers high-fidelity IT/OT simulations for Industrial Control Systems (ICS) environments, though its technical implementations remain proprietary. In the commercial sphere, \textbf{Cyberbit}~\cite{Beuran2025} features realistic environments mapped to the MITRE ATT\&CK framework but requires significant investment and remains closed-source. Domain-specific platforms like the \textbf{Healthcare Cyber Range (HCCR)}~\cite{Beuran2025} and \textbf{SPIDER}~\cite{Beuran2025} offer high-fidelity niche environments (e.g., medical devices, 5G security) but lack broader architectural modularity.

Bader et al. \cite{BADER2026100845} introduced PowerRange, a high-fidelity power-grid CR that integrates hardware-in-the-loop, realistic control-room interaction, and automated scenario derivation via PowerOwl\cite{BADER2026100845}, combining ICT emulation, OT communication, and power simulation. In contrast, the proposed CR is a lightweight, modular smart-energy CR featuring YAML-driven orchestration, QEMU/KVM virtualisation, native Modbus/TCP and MQTT simulation, and a Red–Blue–Green Team framework for technical and human-centric assessment.

As shown in Table~\ref{tab:cyber-range-comparison}, existing open-source platforms typically lack native OT/ICS support, while operational, high-fidelity ranges are rarely open or modular. The proposed CR bridges this gap as a lightweight, modular platform providing native IT/OT protocol simulation (Modbus/TCP, MQTT) and automated YAML-driven orchestration on commodity hardware. Critically, it is the only platform in this comparison to explicitly integrate a tri-team framework (Red, Blue, and \textbf{Green Team}) to address both technical training and organisational readiness.

\begin{table*}[t]
\centering
\small
\caption{Comparative Analysis of Selected CRs and Platforms}
\label{tab:cyber-range-comparison}
\begin{tabularx}{\textwidth}{@{} l c c c c c c @{}}
\toprule
\textbf{Range} & 
\makecell{\textbf{Openness \&}\\\textbf{Automation}} & 
\makecell{\textbf{IT/OT \&}\\\textbf{ICS Support}} & 
\makecell{\textbf{Monitoring \&}\\\textbf{Threat Realism}} & 
\makecell{\textbf{Domain-Specific}\\\textbf{Testbed}} & 
\makecell{\textbf{Academic /}\\\textbf{Training}} & 
\makecell{\textbf{Operational /}\\\textbf{Specialised}} \\
\midrule
\textbf{KYPO}~\cite{9637180} & \cmark & -- & \cmark & -- & \cmark & -- \\
\textbf{AIT}~\cite{10.1145/3424954.3424959} & -- & \cmark & -- & \cmark & -- & \cmark \\
\textbf{CyTrONE}~\cite{10.1145/3011077.3011087} & \cmark & -- & -- & -- & \cmark & -- \\
\textbf{Cyberbit}~\cite{Beuran2025} & -- & -- & \cmark & -- & \cmark & \cmark \\
\textbf{HCCR}~\cite{Beuran2025} & -- & -- & -- & \cmark & -- & \cmark \\
\textbf{SPIDER}~\cite{Beuran2025} & -- & -- & -- & \cmark & -- & \cmark \\
\textbf{PowerRange}~\cite{BADER2026100845} & -- & \checkmark & \checkmark & Smart Energy/Power Grid & \checkmark & -- \\
\textbf{Proposed CR} & \cmark & \cmark & \cmark & Smart Energy/Power Grid & \cmark & -- \\
\bottomrule
\multicolumn{7}{@{}p{\textwidth}@{}}{\footnotesize \textbf{Notes:} \textit{Openness/Automation} includes open-source, IaC, or YAML config. \textit{Monitoring/Threat Realism} covers MITRE ATT\&CK alignment or team exercises. \textit{Domain-Specific} denotes sector-tailored environments (e.g., healthcare, 5G, energy).} \\
\end{tabularx}
\end{table*}

\section{System Architecture and Scenario Design}
\label{sec:Architecture}

\subsection{Platform Architecture and Design Rationale}

\begin{figure}[h!]
    \centering
    \includegraphics[width=\columnwidth]{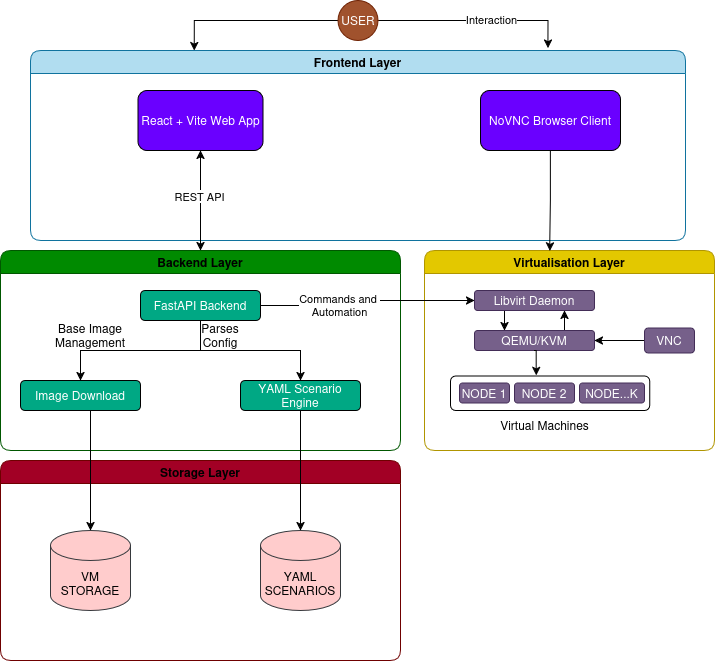}
    \caption{System architecture and component interactions of the cyber range platform, illustrating the workflow from the user-facing frontend to the hypervisor-level virtualisation layer.}
    \label{fig:cyberrange_arch}
\end{figure}

The internal architecture of the proposed CR follows a modular, tiered approach for scalable virtualisation and seamless interaction, as illustrated in Fig.~\ref{fig:cyberrange_arch}. The system comprises four primary layers:
\begin{itemize}
    \item \textbf{Frontend Layer:} A React-based web application providing a dashboard for topology management. A NoVNC client enables direct console access via WebSockets.
    \item \textbf{Backend Layer:} A FastAPI framework core that manages business logic via REST APIs, including an image extraction service and a YAML engine for automated deployment.
    \item \textbf{Storage Layer:} A persistent layer for Virtual Machine (VM) images (ISOs) and YAML-based scenario definitions.
    \item \textbf{Virtualisation Host:} Utilises the Libvirt daemon and QEMU/KVM to manage VM lifecycles. Nodes are exposed through VNC/WebSockets for remote trainee access.
\end{itemize}

\subsection{Design Rationale and Key Differentiators}
Our architectural choices are directly motivated by the limitations of the existing platforms identified in Section~\ref{sec:RelatedWork}. Unlike monolithic or resource-intensive platforms, the proposed CR enables lightweight, automated deployment on commodity hardware. Its FastAPI-based backend and YAML-driven orchestration reduce the barrier to entry for educators compared to closed-source alternatives. This infrastructure-as-code approach provides version-controlled, reproducible, and shareable scenarios that are often absent from operational ranges.

Although containerization offers lower overhead, QEMU/KVM was selected to provide high-fidelity kernel-level isolation, realistic malware execution environments, and native network-stack manipulation required for accurate OT emulation, which container runtime cannot safely replicate.

Furthermore, while platforms like KYPO excel at IT automation, they lack native support for OT protocols. Our explicit layering of protocol simulation such as \textbf{Modbus/TCP and MQTT} is a core differentiator. By co-locating these protocol simulators within the same virtualised environment as standard IT services, we enable realistic \textbf{cascading attack chains} from IT (e.g. WiFi compromise) to OT (e.g., Modbus command injection). This directly contrasts with siloed IT and OT training environments.

\begin{figure}[t]
    \centering
    \includegraphics[width=0.95\columnwidth]{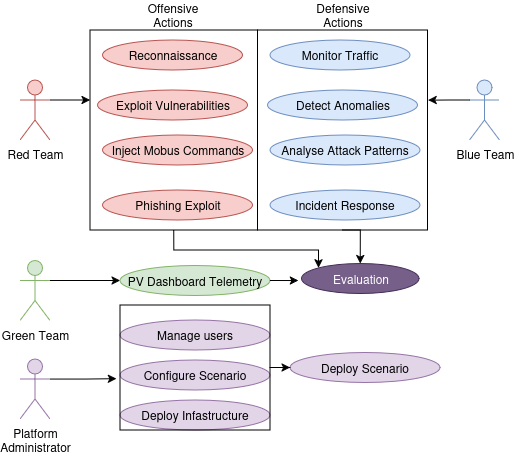}
    \vspace{-2mm}
    \caption{Use-case diagram showing interaction of Red, Blue, and Green Teams}
    \label{fig:usecase}
    \vspace{-4mm}
\end{figure}

Finally, the architecture's support for distinct Red, Blue, and Green Team interfaces is a key design choice for human-centric assessment. As illustrated in Figure \ref{fig:usecase} which enables the structured evaluation not only of technical exploits but also of human factors like operator situational awareness and decision-making under stress, moving beyond the purely technical focus of most existing ranges.

\subsection{Scalability and Resource Management}
The platform is designed with a strong emphasis on minimal resource consumption. Its core system runs as a set of containerised services, including a FastAPI backend, React frontend, and a noVNC browser-based Virtual Network Computing (VNC) proxy, while Virtual Machines (VMs) are orchestrated by the host's native KVM hypervisor. Individual scenarios allocate CPU and memory resources on demand through YAML definitions (e.g., cpu: 2, ram: 4096), controlable based on needs. Entire training topologies can be started and stopped as needed, avoiding the need for large, continuously running infrastructure.

\subsection{Scenario Overview: The PV Plant Attack}
The platform is used to deploy complex scenarios and topologies for simulation and training. While validated here via the PV Plant topology, the underlying YAML orchestration engine is completely decoupled from this use case. It parses arbitrary node definitions, networking configurations, and vulnerability states, allowing administrators to instantiate completely different IT/OT topologies without modifying the platform core.

Can be modeled into segments of civilian and military ICT/OT infrastructures, including Smart Energy Grids comprising Supervisory Control and Data Acquisition (SCADA) systems with Remote Terminal Units (RTUs), Communication Networks (5G for IoT), and Military Systems such as Command and Control (C2) systems. The core scenario is the \textbf{PV Plant} attack chain, set in a fictional smart city, which models a cascading attack from a civilian residence to a military camp. The scenario is designed to involve all three teams, as shown in Fig.~\ref{fig:usecase}. The complete cascading cyber-attack chain is illustrated in Figure \ref{fig:attackchain}.

\subsection{Attack Vector Decomposition and MITRE ATT\&CK Mapping}  \label{sec:attackchain}
To provide a structured and realistic training environment, the attack vectors in the PV Plant scenario are decomposed according to standardised adversary behaviour models. Table~\ref{tab:attack-decomposition} details the specific vulnerabilities, exploits, and MITRE ATT\&CK mappings for each level. This decomposition allows the Red Team to follow a realistic, multi-stage attack path and enables the Blue Team to correlate alerts to specific tactics and techniques.

\begin{table}[htbp]
\centering
\caption{Decomposition of Attack Vectors with MITRE ATT\&CK Mapping}
\label{tab:attack-decomposition}
\footnotesize 
\setlength{\tabcolsep}{3pt} 
\begin{tabularx}{\linewidth}{@{} l >{\raggedright\arraybackslash}p{1.8cm} X >{\raggedright\arraybackslash}p{2.2cm} @{}}
\toprule
\textbf{Level} & \textbf{Initial Vulnerability} & \textbf{Attack Path / Exploit} & \textbf{MITRE ATT\&CK Tactic (ID)} \\
\midrule
\begin{tabular}[t]{@{}l@{}}L1: \\ Smart \\ Home\end{tabular} & 
Weak WiFi password & 
1. Reconnaissance (scanning for open ports on PV inverter) \newline 
2. Exploitation (Modbus command injection) & 
Initial Access (TA0001) \newline 
Command \& Control (TA0011) \\
\midrule
\begin{tabular}[t]{@{}l@{}}L2: \\ Grid \\ Disr\end{tabular} & 
Unencrypted MQTT traffic \& Default credentials & 
\textbf{Path A (MQTT Hijack):} Intercept and inject fake power telemetry & 
Collection (TA0009) \newline 
Impact (TA0040) \\
\cmidrule{2-4}
 & 
Weak API rate limiting \& logging & 
\textbf{Path B (API Hijack):} Credential stuffing on REST API to reduce grid allocation & 
Credential Access (TA0006) \newline 
Defense Evasion (TA0005) \\
\cmidrule{2-4}
 & 
MQTT broker misconfig. (anon. publish) & 
\textbf{Path C (Dictionary):} Brute-force MQTT topic write access to alter commands & 
Privilege Esc. (TA0004) \newline 
Lateral Mvmt. (TA0008) \\
\bottomrule
\end{tabularx}
\end{table}

\subsection{Full Attack Path Description}
\begin{itemize}
    \item \textbf{Level 1: Smart Home Attack (Unauthorised shutdown of a residential PV system).}
    \begin{itemize}
        \item \textit{Initial Access:} The Red Team conducts network reconnaissance to identify and gain unauthorised access to the residence’s Wi-Fi network by cracking a weak Wi-Fi Protected Access 2 Pre-Shared Key (WPA2-PSK) password.
        \item \textit{Exploitation:} From the compromised network, the Red Team scans for the PV inverter, identifies an open Modbus Transmission Control Protocol (Modbus/TCP) port (502), and sends a malicious write-coil command to halt energy production. The Green Team observes a sudden drop in power output to zero on its dashboard.
    \end{itemize}
    \item \textbf{Level 2: Smart Grid Disruption (Causing a power shortage in a military area).}
    \begin{itemize}
        \item \textit{Initial Access (Alternative Paths):} Using the residential compromise as a pivot, the Red Team attacks the central IoT platform via three vectors:
        \begin{enumerate}
            \item \textbf{MQTT Session Hijacking:} Intercepting unencrypted MQTT traffic to manipulate power distribution data.
            \item \textbf{REST API Hijacking:} Using credentials obtained from a phishing attack (injected by the platform) to command the grid to reduce power allocation directly.
            \item \textbf{Dictionary Attack:} Cracking weak packet encryption keys for the MQTT broker to inject malicious commands.
        \end{enumerate}
        \item \textit{Impact:} Successful exploitation leads to a significant, targeted power shortage in the military area. The Blue Team must correlate alerts (e.g., failed MQTT auth, API rate violations) across the network, application, and ICS layers to identify the attack.
    \end{itemize}
\end{itemize}

\begin{figure}[t]
    \centering
    \includegraphics[width=\linewidth]{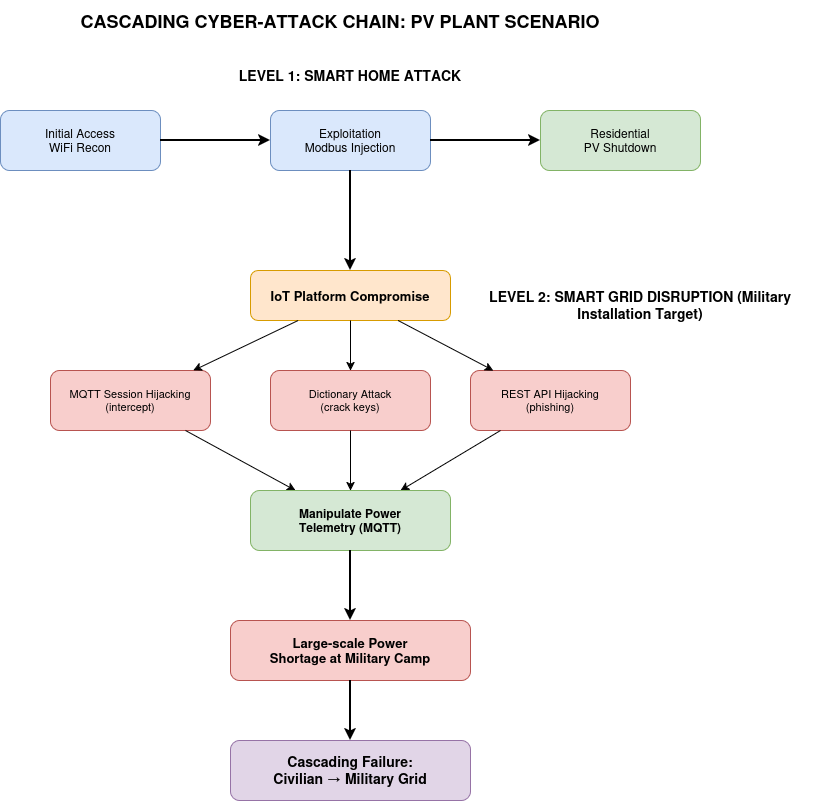}
    \caption{Cyber attack chains in a smart energy systems, showing the progression from residential compromise to grid disruption for the proposed scenario.}
    \label{fig:attackchain}
\end{figure}

\subsection{Expanded Smart Energy Testbed Topology}
\label{sec:expanded_topology}

\begin{figure}[h!]
    \centering
    \includegraphics[width=\columnwidth]{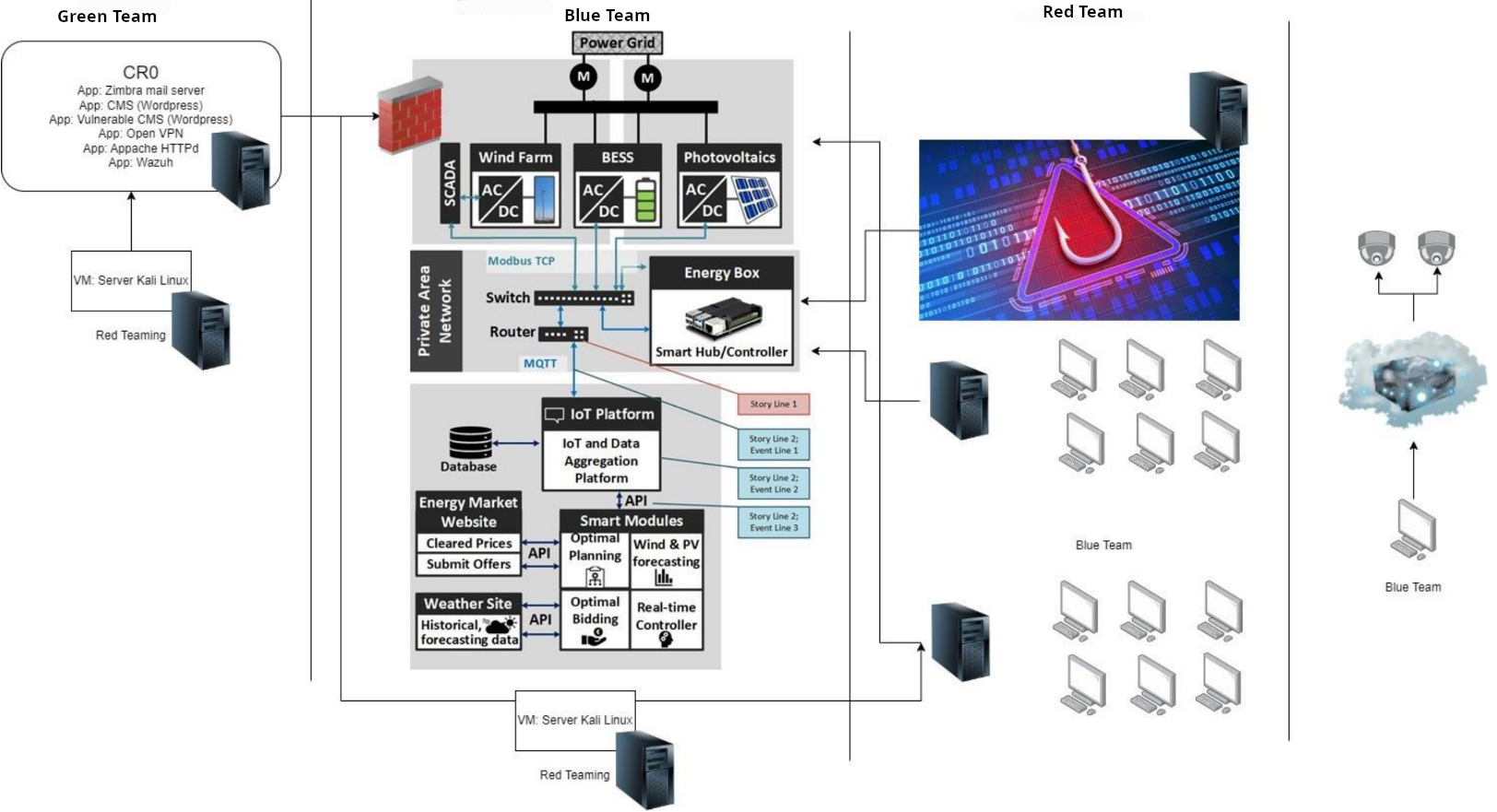}
    \caption{Expanded Smart Energy Testbed: IT, OT, market APIs, and team vantage points.}
    \label{fig:expanded_topology}
\end{figure}

Figure~\ref{fig:expanded_topology} presents the complete testbed for the PV Plant scenario, extending beyond the residential inverter to include a wind farm, a Battery Energy Storage System (BESS) using Modbus/TCP, energy-market and weather Application Programming Interfaces (APIs), and Information Technology (IT) services such as Zimbra, WordPress, OpenVPN, and Wazuh. Two Kali Linux Virtual Machines (VMs) support Red Team operations, while the Blue Team monitors network traffic through the switch. This configuration enables IT$\rightarrow$OT$\rightarrow$market attack chains, such as compromising WordPress to manipulate market APIs and inject Modbus commands into generation assets, while the Green Team observes physical and economic anomalies through operational dashboards. These capabilities are further described by Christou et al.~\cite{christou2026actingplatformcyberranges}. 

\section{Evaluation}
\label{sec:Evaluation}

As depicted in Figure \ref{fig:attackchain}, the proposed CR simulates industrial and IoT protocols essential to the PV Plant cascading attack scenario. For \textbf{Level 1}, the range deploys virtualised inverters that communicate via Modbus/TCP, a standard ICS protocol that, when unsecured, is vulnerable to various forms of exploitation. Trainees (Red Team) can conduct network reconnaissance and execute unauthorised commands. Blue Team monitors Modbus traffic for anomalies (e.g., function code violations or write-coil attacks).

For \textbf{Level 2}, the proposed CR uses an MQTT broker architecture that simulates a dense network of sensors and controllers exchanging power telemetry data. By intentionally simulating these exchanges over unsecured channels, the platform facilitates training in MQTT session hijacking and traffic interception. These vectors enable participants to manipulate energy output data and observe the resulting power shortages.

\subsection{Red Team Design}

The Red Team acts as the adversary, executing the attack chains described in Section \ref{sec:attackchain}. CR supports Red Team operations through:

\begin{itemize}
    \item \textbf{Scenario injection points:} The YAML orchestration engine pre-configures vulnerabilities (e.g., default credentials, open MQTT ports, weak API authentication) that the Red Team can discover and exploit.
    \item \textbf{Protocol exploitation tools:} The platform includes virtualised instances of common penetration testing tools (e.g., \texttt{modbus-cli}, MQTT explorers, \texttt{curl} for REST API) accessible from dedicated attacker VMs.
    \item \textbf{Multi-vector attack execution:} Red Team members can simultaneously launch MQTT session hijacking, dictionary attacks against MQTT write-keys, and phishing campaigns targeting Green Team dashboards.
    \item \textbf{Stealth and persistence:} The platform logs all Red Team actions for after-action review but does not artificially constrain them, allowing trainees to practice evasion techniques.
\end{itemize}

The Red Team's primary goal is to achieve the cascading failure (residential PV shutdown followed by military grid disruption) while avoiding detection by the Blue Team.

\subsection{Blue Team Design}
The Blue Team monitors, detects, and responds to ongoing threats, with all network traffic routed through their vantage point for full-visibility surveillance. The platform provides integrated defensive capabilities including: 

\begin{itemize}
    \item \textbf{Centralised logging and alerting}: Ntopng aggregates logs from all virtualised assets, covering Modbus queries, MQTT messages, API calls, and system authentication attempts
    \item \textbf{Network traffic analysis}: Packet capture (PCAP) archives are available across all network segments, allowing defenders to use built-in tools like Zeek and Wireshark to investigate suspicious flows.
    \item  \textbf{Anomaly detection}: Ntopng and PiHole serve as detection mechanisms to highlight Modbus function code violations, MQTT authentication failures, REST API rate limits, and unauthorized access to operational dashboards
\end{itemize}

Blue Team performance is evaluated based on detection latency, alert accuracy, response effectiveness, and communication with the Green Team.

\subsection{Green Team Framework}

In the proposed CR, the \textbf{Green Team} is designed to assume the operational role of PV plant engineers, serving as the primary monitoring entity for the energy infrastructure. This design enables future assessment of human operators' defensive posture and situational awareness. The framework includes:

\begin{itemize}
    \item \textbf{Operational Monitoring Dashboard:} A functional PV dashboard visualising real-time telemetry (inverter voltage, power output, MQTT broker status). Green Team participants monitor these data streams to distinguish routine fluctuations from malicious activity.
    \item \textbf{Social Engineering Vector Simulation:} The environment injects deceptive communications mimicking administrative or maintenance requests. Participants are tasked with identifying attempts at credential theft and unauthorised access.
\end{itemize}

This pedagogical design moves beyond technical validation to enable analysis of the "human element" in critical infrastructure protection~\cite{10.1007/978-3-031-94159-7_14}, though empirical validation remains future work. Together, Red, Blue, and Green Teams provide a complete evaluation ecosystem: Red challenges defences, Blue measures detection and response, Green assesses operational awareness.

\section{Conclusions and Future Directions}
\label{sec:Conclusions_FutureWork}

This paper presented a modular CR platform for simulating smart energy systems and cyber-physical infrastructures. The platform combines a FastAPI backend, KVM-based virtualisation, and YAML-based orchestration to support the automated deployment of complex training environments. It also integrates protocol simulation for Modbus/TCP and MQTT, enabling realistic cascading attack scenarios across interconnected ICT and OT components. Through the representative PV Plant scenario, the paper demonstrated how multi-stage attack chains can propagate from residential IoT environments to wider operational disruptions.

The paper also introduced a Green Team framework for human-centric assessment. To better represent segmented or air-gapped operational environments, future iterations will quantify this assessment via integrated dashboard analytics that log key performance metrics, including detection time($T_{det}$), operator reporting latency, and checklist compliance rates.

By complementing Red and Blue Team activities, this framework aims to support the evaluation of awareness, coordination, decision-making, and response behaviour in cyber-physical training scenarios. This extends the role of the CR beyond technical exploitation and detection, positioning it as a broader environment for assessing cyber readiness.

Transitioning the proposed CR to a production-grade platform involves several key directions. Future work will implement role-based access control for all team personas and enhance industrial network emulation through software-defined networking, Virtual Local Area Network (VLAN) tagging, and multi-homed interfaces to better represent segmented or air-gapped operational environments. The integration of Large Language Models (LLMs) through local inference, such as Ollama, for dynamic scenario generation and real-time guidance will also be explored, alongside cross-domain federation across multiple hardware-in-the-loop deployments and international CRs.

Future work will also include a comprehensive performance benchmark to empirically measure VM provisioning latency, CPU and RAM utilisation under peak training loads, and the hardware overhead introduced by the QEMU/KVM virtualisation layer.

\bibliographystyle{IEEEtran}
\bibliography{csr2026-wshp-kios-cr}

@article{kampourakis2026ontology,
  author  = {Kampourakis, Vyron and Takaronis, Michail and Gkioulos, Vasileios and Katsikas, Sokratis},
  title   = {An Ontology-Based Framework for Semantic Representation of the Cyber Range Domain},
  journal = {Journal of Cybersecurity and Privacy},
  volume  = {6},
  number  = {2},
  pages   = {76},
  year    = {2026},
  date    = {2026-04-21},
  publisher = {MDPI AG},
  doi     = {10.3390/jcp6020076},
  url     = {https://doi.org/10.3390/jcp6020076}
}

@article{kampourakis2024reference,
author = {Kampourakis, Vyron and Gkioulos, Vasileios and Katsikas, Sokratis},
year = {2024},
month = {11},
pages = {103917},
title = {A step-by-step definition of a reference architecture for cyber ranges},
volume = {88},
journal = {Journal of Information Security and Applications},
doi = {10.1016/j.jisa.2024.103917}
}

@article{Kampourakis2025LLMAssistedAF,
  title={LLM-Assisted AHP for Explainable Cyber Range Evaluation},
  author={Vyron Kampourakis and Georgios Kavallieratos and Georgios P. Spathoulas and Vasileios Gkioulos and Sokratis K. Katsikas},
  journal={ArXiv},
  year={2025},
  volume={abs/2512.10487},
  url={https://api.semanticscholar.org/CorpusID:283737288}
}

@article{admass2024cyber,
  title={Cyber security: State of the art, challenges and future directions},
  author={Admass, Wasyihun Sema and Munaye, Yirga Yayeh and Diro, Abebe Abeshu},
  journal={Cyber Security and Applications},
  volume={2},
  pages={100031},
  year={2024},
  publisher={Elsevier}
}

@article{yu2023survey,
  title={A survey on cyber--physical systems security},
  author={Yu, Zhenhua and Gao, Hongxia and Cong, Xuya and Wu, Naiqi and Song, Houbing Herbert},
  journal={IEEE Internet of Things Journal},
  volume={10},
  number={24},
  pages={21670--21686},
  year={2023},
  publisher={IEEE}
}

@article{yamin2020cyber,
  title={Cyber ranges and security testbeds: Scenarios, functions, tools and architecture},
  author={Yamin, Muhammad Mudassar and Katt, Basel and Gkioulos, Vasileios},
  journal={Computers \& Security},
  volume={88},
  pages={101636},
  year={2020},
  publisher={Elsevier}
}

@techreport{ECSO2020,
  author   = {European Cyber Security Organisation (ECSO)},
  title    = {Understanding Cyber Ranges: From Hype to Reality},
  institution = {European Cyber Security Organisation},
  year      = {2020},
  month     = mar,
  note      = {White Paper}
}

@InProceedings{Virag2021,
  author={Virág, Csaba and Čegan, Jakub and Lieskovan, Tomáš and Merialdo, Matteo},
  booktitle={2021 IEEE International Conference on Cyber Security and Resilience (CSR)}, 
  title={The Current State of The Art and Future of European Cyber Range Ecosystem}, 
  year={2021},
  volume={},
  number={},
  pages={390-395},
  doi={10.1109/CSR51186.2021.9527931}}

@Article{Park2022,
AUTHOR = {Park, Moosung and Lee, Hyunjin and Kim, Yonghyun and Kim, Kookjin and Shin, Dongkyoo},
TITLE = {Design and Implementation of Multi-Cyber Range for Cyber Training and Testing},
JOURNAL = {Applied Sciences},
VOLUME = {12},
YEAR = {2022},
NUMBER = {24},
ARTICLE-NUMBER = {12546},
ISSN = {2076-3417},
DOI = {10.3390/app122412546}
}

@article{PRUMMER2025104206,
title = {Assessing the effect of cybersecurity training on End-users: A Meta-analysis},
journal = {Computers \& Security},
volume = {150},
pages = {104206},
year = {2025},
issn = {0167-4048},
doi = {https://doi.org/10.1016/j.cose.2024.104206},
url = {https://www.sciencedirect.com/science/article/pii/S016740482400511X},
author = {Julia Prümmer and Tommy {van Steen} and Bibi {van den Berg}}
}

@InProceedings{10.1007/978-3-030-01168-0_26,
author="Kokkonen, Tero
and Puuska, Samir",
editor="Galinina, Olga
and Andreev, Sergey
and Balandin, Sergey
and Koucheryavy, Yevgeni",
title="Blue Team Communication and Reporting for Enhancing Situational Awareness from White Team Perspective in Cyber Security Exercises",
booktitle="Internet of Things, Smart Spaces, and Next Generation Networks and Systems",
year="2018",
publisher="Springer International Publishing",
address="Cham",
pages="277--288",
isbn="978-3-030-01168-0"
}

@article{YULIANTO2025100077,
title = {Enhancing cybersecurity resilience through advanced red-teaming exercises and MITRE ATT\&CK framework integration: A paradigm shift in cybersecurity assessment},
journal = {Cyber Security and Applications},
volume = {3},
pages = {100077},
year = {2025},
issn = {2772-9184},
doi = {https://doi.org/10.1016/j.csa.2024.100077},
url = {https://www.sciencedirect.com/science/article/pii/S2772918424000432},
author = {Semi Yulianto and Benfano Soewito and Ford Lumban Gaol and Aditya Kurniawan}
}

@misc{christou2026actingplatformcyberranges,
  title={ACTING: A Platform for Cyber Ranges Federation}, 
  author={Christou, Kyriakos and others},
  year={2026},
  eprint={2605.12170},
  archivePrefix={arXiv},
  primaryClass={cs.CR},
  url={https://arxiv.org/abs/2605.12170}
}

@InProceedings{10.1007/978-3-031-94159-7_14,
author="Yiangou, Ioannis
and Stavrou, Eliana",
editor="Stephanidis, Constantine
and Antona, Margherita
and Ntoa, Stavroula
and Salvendy, Gavriel",
title="Developing Green IT and Cybersecurity Competencies Through Virtualization: Upskilling Learners Across Disciplines",
booktitle="HCI International 2025 Posters",
year="2025",
publisher="Springer Nature Switzerland",
address="Cham",
pages="129--139",
isbn="978-3-031-94159-7"
}

@Inbook{Beuran2025,
author="Beuran, Razvan",
title="Cyber Ranges",
bookTitle="Cybersecurity Education and Training",
year="2025",
publisher="Springer Nature Singapore",
address="Singapore",
pages="221--259",
isbn="978-981-96-0555-2",
doi="10.1007/978-981-96-0555-2_11",
url="https://doi.org/10.1007/978-981-96-0555-2_11"
}

@inproceedings{10.1145/3011077.3011087,
author = {Pham, Cuong and Tang, Dat and Chinen, Ken-ichi and Beuran, Razvan},
title = {CyRIS: a cyber range instantiation system for facilitating security training},
year = {2016},
isbn = {9781450348157},
publisher = {Association for Computing Machinery},
address = {New York, NY, USA},
url = {https://doi.org/10.1145/3011077.3011087},
doi = {10.1145/3011077.3011087},
booktitle = {Proceedings of the 7th Symposium on Information and Communication Technology},
pages = {251–258},
numpages = {8},
location = {Ho Chi Minh City, Vietnam},
series = {SoICT '16}
}

@inproceedings{10.1145/3424954.3424959,
author = {Leitner, Maria and Frank, Maximilian and Hotwagner, Wolfgang and Langner, Gregor and Maurhart, Oliver and Pahi, Timea and Reuter, Lenhard and Skopik, Florian and Smith, Paul and Warum, Manuel},
title = {AIT Cyber Range: Flexible Cyber Security Environment for Exercises, Training and Research},
year = {2021},
isbn = {9781450375993},
publisher = {Association for Computing Machinery},
address = {New York, NY, USA},
url = {https://doi.org/10.1145/3424954.3424959},
doi = {10.1145/3424954.3424959},
booktitle = {Proceedings of the 2020 European Interdisciplinary Cybersecurity Conference},
articleno = {2},
numpages = {6},
location = {Rennes, France},
series = {EICC '20}
}

@INPROCEEDINGS{9637180,
  author={Vykopal, Jan and Čeleda, Pavel and Seda, Pavel and Švábenský, Valdemar and Tovarňák, Daniel},
  booktitle={2021 IEEE Frontiers in Education Conference (FIE)}, 
  title={Scalable Learning Environments for Teaching Cybersecurity Hands-on}, 
  year={2021},
  volume={},
  number={},
  pages={1-9},
  doi={10.1109/FIE49875.2021.9637180}}

@article{BADER2026100845,
    title = {PowerRange: An immersive cyber range for power grid operators},
    journal = {International Journal of Critical Infrastructure Protection},
    volume = {53},
    pages = {100845},
    year = {2026},
    issn = {1874-5482},
    doi = {https://doi.org/10.1016/j.ijcip.2026.100845},
    url = {https://www.sciencedirect.com/science/article/pii/S187454822600017X},
    author = {Lennart Bader and Eric Wagner and Martin Serror}
}

\end{document}